\documentclass[aps,prl,preprint,groupedaddress]{revtex4}
\def\subrightarrow#1{%                          % #1 under arrow
  \setbox0=\hbox{%                              % set a box
    $\displaystyle\mathop{}%                    % no mathop
    \limits_{#1}$}%                             % just limits
  \dimen0=\wd0%                                 % get width
  \advance \dimen0 by .5em%                     % add a bit
  \mathrel{%                                    % space like =
    \mathop{\hbox to \dimen0{\rightarrowfill}}% % arrow to width
       \limits_{#1}}}                           % text below

\begin{document}

%\preprint{}
\title{Improved Calculation of Electroweak Radiative Corrections and
the Value of $V_{ud}$}


\author{William J. Marciano $^{1}$ and Alberto Sirlin $^{2}$} 
%\thanks{}
\affiliation{$^{1}$ Brookhaven National Laboratory, Upton, NY 11973\\
  $^{2}$ New York University, Department of Physics, 4 Washington
  Place, New York, NY 10003}


\date{September 2005}

\begin{abstract}
A new method for computing hadronic effects on electroweak radiative
corrections to low-energy weak interaction semileptonic processes is
described. It employs high order perturbative QCD results originally
derived for the Bjorken sum rule along with a large N QCD-motivated
interpolating function that matches long and short-distance loop
contributions. Applying this approach to the extraction of the CKM
matrix element $V_{ud}$ from superallowed nuclear beta decays
reduces the theoretical loop uncertainty by about a factor of 2 and
gives $V_{ud} = 0.97377(11)(15)(19)$. Implications for CKM unitarity
are briefly discussed.
\end{abstract}

%\pacs{}
%\keywords{}

\maketitle

Precision studies of low-energy semileptonic weak-charged and neutral
current processes can be used to test the $SU(3)_{c} \times SU(2)_{L}
\times U(1)_{Y}$ Standard Model at the quantum loop level and probe
for potential ``new physics'' effects.  Examples for which a fraction
of a percent experimental sensitivity has already been achieved
include: pion, neutron and nuclear beta decays \cite{towner1}, as well
as atomic parity violation \cite{bennett}. In those cases, electroweak
radiative corrections (RC) have been computed \cite{sirlin1, marciano1,
  marciano2} and found to be significant (of order several
percent). They must be included in any meaningful confrontation
between theory and experiment.

Of course, inherent to any low-energy semileptonic process are
uncertainties due to strong interactions, since quarks are
involved. To minimize such effects, one often focuses on weak vector
current-induced reactions, where CVC (conserved vector current)
protects those amplitudes at tree level from strong interaction
corrections in the limit of zero momentum transfer. However, even for
those amplitudes, electroweak loop corrections can involve weak
axial-vector effects not protected by CVC, which give rise to
hadronic (strong interaction) uncertainties in their evaluation
\cite{sirlin1, marciano1}. In this paper, we focus on the best known
and tested examples of that phenomenon, the electroweak radiative
corrections to neutron and correspondingly superallowed nuclear beta
decays along with their implications for the extraction of the CKM matrix element
$V_{ud}$. However, the method we describe is quite general and can be
easily applied to other charged and neutral current semileptonic
low-energy reactions.

The extraction of $V_{ud}$ (in fact all CKM matrix elements) entails
normalizing a semileptonic reaction rate with respect to the muon
lifetime, or equivalently the Fermi constant derived from it
\begin{equation}
G_{\mu} = 1.16637(1) \times 10^{-5} GeV^{-2}
\end{equation}
\noindent For high precision, electroweak radiative corrections to
both processes must be included and hadronic as well as environmental
effects (e.g., nuclear structure) must be controlled. Toward that end,
super-allowed ($0^{+} \rightarrow 0^{+}$) nuclear beta decay
transitions are very special since they only involve the weak vector
current at tree level. Small violations of CVC due to the up-down mass
difference or non-zero momentum transfer are small $\sim {\cal
  O}(10^{-5})$ and can generally be neglected (or incorporated). Such an
analysis leads to the very accurate relationship \cite{czarnecki, hardy}
\begin{equation}
|V_{ud}|^{2} = \frac {2984.48(5) {\rm sec}} {ft(1+RC)} \
({\rm Superallowed \ \beta-decays})
\end{equation}
\noindent where ft is the product of a phase space statistical decay
rate factor f (which depends on the Q value of a specific nuclear
beta decay) and its measured half-life t. RC designates the total
effect of all radiative corrections relative to muon decay as well as
QED-induced nuclear structure isospin violating effects. It is nucleus
dependent, ranging from about $+3.1 \%$ to $+3.6 \%$ for the nine
best-measured super-allowed decays. So, measuring Q and t combined
with computing RC determines $V_{ud}$. A similar formula will be given
later for neutron beta decay. In that case, the Q value =
$m_{n}-m_{p}$ is very precisely known, but in addition to the neutron
lifetime, $g_{A} \equiv G_{A}/G_{V}$  must be accurately
measured because both weak axial and vector currents contribute at
tree level \cite{towner1, czarnecki}.

Our main goal in this paper is to reduce the hadronic uncertainty
in the radiative corrections to super-allowed nuclear beta decays and
thereby improve the determination of $V_{ud}$. The need for such an
improvement is well illustrated by a survey of ft values and RC for
super-allowed beta decays by Hardy and Towner \cite{hardy}, more
recently updated by G. Savard et al. \cite{savard}, which found
\begin{equation}
V_{ud}=0.9736(2)(4)_{EW}
\end{equation}
\noindent where the first uncertainty stems primarily from nuclear
structure corrections (including ${\cal O}(Z^{2}\alpha^{3})$ effects)
and very small ft value errors while the second, dominant error is due
to hadronic uncertainties in electroweak loop effects. Although, as we
mention later, the first error may currently be an underestimate and
the central value of $V_{ud}$ could shift due to future Q value
updates, it is clear that the hadronic loop uncertainty, which comes
from weak axial-current loop effects, currently limits the
determination of $V_{ud}$ and must be improved if further progress is
to be made.

Here, we describe a new method for controlling hadronic uncertainties
in the radiative corrections to neutron and super-allowed nuclear beta decays.
It validates our previous results \cite{marciano1, czarnecki} increasing
$V_{ud}$ by only a small +0.00007, but reduces the loop uncertainty by
about a factor of 2, $(0.0004)_{EW} \rightarrow (0.0002)_{EW}$ as we
now demonstrate.

The one-loop electroweak radiative corrections to the neutron (vector
current contribution) and super-allowed nuclear beta decays are given
by \cite{sirlin1, marciano1, sirlin2}
\begin{equation}
RC_{EW} = \frac {\alpha}{2\pi} \left\{ \overline{g}(E_{m})+3ln
\frac{m_Z}{m_p} + ln \frac{m_Z}{m_A} + A_{g} + 2C_{Born} \right\}  
\end{equation}
\noindent The first two terms result from loop corrections and
bremsstrahlung involving electromagnetic and weak vector current
interactions, with $\overline{g} (E_{m})$ a universal function
\cite{sirlin2} integrated over phase space and $3ln \frac{m_Z}{m_p}$ a
short-distance loop effect. They are not affected by strong
interactions up to ${\cal O}(\frac{\alpha}{\pi}
\frac{E_{m}}{m_{p}})\simeq 10^{-5}$ corrections which can be neglected
at our present level of accuracy. Higher order leading logs of order
$\alpha^{n}ln^{n}(m_{Z}/m_{p})$ etc. can be summed via a
renormalization group analysis \cite{marciano1} and ${\cal
O}(Z\alpha^{2})$ as well as ${\cal O}(Z^{2}\alpha^{3})$ contributions
have been computed for high Z nuclei \cite{sirlin3}. They
will not be explicitly discussed here, but are included in our final
results.

The last three terms in eq. (4) are induced by weak axial-vector
current loop effects. Their primary source is the $\gamma W$ box
diagram which involves the time-ordered product of the electromagnetic
and weak axial-vector currents. That product contains a leading vector
current component which contributes to $0^{+}\rightarrow0^{+}$ nuclear
transition elements. Employing the current algebra formulation, one
finds \cite{sirlin1}
\begin{equation}
Box (\gamma W)_{VA}= \frac{\alpha}{8\pi} \int\limits_{0}^{\infty}
dQ^{2} \ \frac{m_{W}^{2}}{Q^{2}+m_{W}^{2}} F(Q^{2})
\end{equation}
\noindent where Q is a Euclidean loop momentum integration variable.

Previous estimates of eq. (5) employed the operator product expansion
plus lowest order QCD correction to obtain the leading effect
\cite{sirlin1, marciano1}
\begin{equation}
F(Q^{2}) \subrightarrow{Q^{2}\rightarrow \infty} \frac{1}{Q^{2}}
\left[ 1- \frac {\alpha_{s}(Q^{2})}{\pi} \right] + {\cal
  O}(\frac{1}{Q^{4}})
\end{equation}
\noindent Integrating over the range $m^{2}_{A}\leq Q^{2} < \infty $
and combining with smaller vertex corrections
and ZW box diagrams, that prescription gave a short-distance amplitude 
contribution
\begin{equation}
\frac {\alpha}{4\pi} \left[ln \frac{m_{Z}}{m_{A}} + A_{g} \right],\ A_{g} \simeq -0.34 \nonumber
\end{equation}
\noindent In the numerical estimate, the low energy cutoff was chosen
to be $m_{A}=1.2 GeV$, roughly the mass of the $A_{1}$ resonance, and
the error was estimated by allowing $m_{A}$ to vary up or down by a
factor of 2. Such a heuristic, albeit crude procedure led to a
$\pm0.0004$ uncertainty in $V_{ud}$. For the long-distance $\gamma W$
box diagram contribution, nucleon electromagnetic and axial-vector dipole form
factors were used to find for neutron decay \cite{marciano1,
marciano2}
\begin{equation}
C_{Born}(\rm {neutron}) \simeq 0.8g_{A}(\mu_{n}+\mu_{p}) \simeq 0.89
\end{equation}
\noindent where $g_{A}\simeq 1.27$ and $\mu_{n}+\mu_{p}=0.88$ is the nucleon isoscalar
magnetic moment. In the case of superallowed nuclear decays, nuclear
quenching modifies $C_{Born}$ (neutron) and nucleon-nucleon
electromagnetic effects must be included \cite{towner2}. Overall, in the
case of a neutron, axial-vector-induced one-loop RC to the decay rate amount
to $0.67(8) \%$. Roughly the same uncertainty $\pm 0.08 \%$ applies to
superallowed nuclear decays.

To reduce the hadronic uncertainty in RC, we have carried out a new
analysis of the $\gamma W$ box diagram axial-vector-induced radiative
corrections that incorporates the following $F(Q^{2})$ improvements
\cite{marciano3}:\\

1) Short Distances $(1.5 GeV)^{2} \leq Q^{2} < \infty$, a domain where
   QCD corrections remain perturbative.
\begin{equation}
F(Q^{2}) = \frac{1}{Q^{2}} \left[ 1 - \frac {\alpha_{s}(Q^{2})_{\overline{MS}}}{\pi} -
  C_{2}(\frac {\alpha_{s}(Q^{2})_{\overline{MS}}}{\pi})^{2} - C_{3} (
  \frac{\alpha_{s}(Q^{2})_{\overline{MS}}}{\pi})^{3} \right]
\end{equation}
\begin{eqnarray}
C_{2} & = & 4.583-0.333N_{F} \\
C_{3} & = & 41.440-7.607N_{F}+0.177N_{F}^{2}
\end{eqnarray}
\noindent where $N_{F}$ = number of effective quark flavors. \\

\noindent 2) Intermediate Distances $((0.823GeV)^{2} \leq Q^{2} <
(1.5GeV)^{2})$
\begin{equation}
F(Q^{2})= \frac{-1.490}{Q^{2}+m_{\rho}^{2}} + \frac{6.855}{Q^{2} +
  m_{A}^{2}} - \frac{4.414}{Q^{2} + m_{\rho'}^{2}}
\end{equation}

\begin{eqnarray}
m_{\rho} & = & 0.776 GeV\\
m_{A} & = & 1.230 GeV\\
m_{\rho'} & = & 1.465 GeV
\end{eqnarray}

\noindent 3) Long Distances: $0 \leq Q^{2} \leq (0.823 GeV)^{2}$

Integrating the long-distance amplitude up to $Q^{2} = (0.823 GeV)^{2}$,
where the integrand matches the interpolating function, and using an
update of the nucleon electromagnetic and axial-current dipole form
factors, we find
\begin{equation}
C_{Born}({\rm neutron})\simeq 0.829
\end{equation}
\noindent a reduction from our own previous result in eq. (8), where
the integration was carried up to $Q^{2}= \infty$.

Details of the above calculations will be given in a subsequent
publication \cite{marciano3}. Here, we briefly discuss the results of
the above analysis  and its implications.

The QCD corrections to the asymptotic form of $F(Q^{2})$ have been
given in eq. (9) to ${\cal O}(\alpha^{3}_{s})$. The additional terms
are identical (in the chiral limit) to QCD corrections to the Bjorken
sum rule \cite{bjorken} for polarized electroproduction and can be
read off from well-studied calculations \cite{beg, gorishny} for that
process. Their validity has been well tested experimentally \cite{deur}.

The interpolating function in eq. (12) is motivated by large N QCD
which predicts it should correspond to an infinite sum of vector and
axial-vector resonances \cite{witten}. We impose three conditions that
determine the residues: i) The integral of eqs. (5) and (12) should
equal that of eqs. (5) and (9) in the asymptotic domain $(1.5
GeV)^{2} \leq Q^{2} \leq \infty$, which amounts to a matching
requirement between domains 1 and 2, ii) In the large $Q^{2}$ limit,
the coefficient of the $1/Q^{4}$ term in the expansion of eq. (12)
should vanish as required by chiral symmetry \cite{small}, iii) The
interpolator should vanish at $Q^{2}=0$ as required by chiral
perturbation theory. Three conditions limit us to three resonances.

The $Q^{2}=(0.823GeV)^{2}$ match between domains 2 and 3 was chosen to
be the value at which eq. (12) equals the integrand of the
long-distance contribution. Interestingly, that matching
occurs near the $\rho$ mass. A novel technical point in the
formulation is that in the evaluation of the Feynman diagrams
associated with the long-distance contributions the integral over the
auxiliary variables is carried out first. This leads to integrands
that depend on $Q^{2}$ and can therefore be matched with eq. (12).

Using this approach, we find that at the one-loop electroweak level
the last three terms in eq. (4) are effectively replaced by 2.82
$\frac {\alpha}{\pi}$ in the case of neutron decay. Comparison with
eqs. (4) and (8) in conjunction with $m_{A}=1.2 GeV$, $A_{g}=-0.34$
shows that in the new formulation these corrections are reduced by
$1.4 \times 10^{-4}$, which increases $V_{ud}$ by $7 \times
10^{-5}$. The smallness of that shift is a validation of our previous
result \cite{marciano1, czarnecki}.

More important than the small reduction in the radiative corrections,
our new method provides a more systematic estimate of the hadronic
uncertainties as well as experimental verification of its validity
\cite{deur}. Allowing for a $\pm 10 \%$ uncertainty for the $C_{Born}$
correction in eq. (16), a $\pm 100 \%$ uncertainty for the
interpolator contribution in the $(0.823GeV)^{2} \leq Q^{2} <
(1.5GeV)^{2}$ region and $\pm 0.0001$ uncertainty from neglected
higher order effects, we find the total uncertainty in the electroweak
radiative corrections is $\simeq \ \pm 0.00038$ which leads to
$\simeq$ a $\pm 0.00019$ uncertainty in $V_{ud}$. That corresponds to more than a
factor of 2 reduction in the loop uncertainty from hadronic effects.

Employing our new analysis, we find the improved relationship between
$V_{ud}$, the neutron lifetime and $g_{A} \equiv G_{V}/G_{A}$
\begin{equation}
|V_{ud}|^{2}= \frac {4908.7(1.9)sec}{\tau_{n} (1+3g_{A}^{2})} \ (\rm
 {neutron})
\end{equation}
\noindent Future precision measurements of $\tau _{n}$ and $g_{A}$
used in conjunction with eq. (17) will ultimately be the best way to
determine $V_{ud}$, but for now it is not competitive
\cite{czarnecki}.

In the case of superallowed ($0^{+}\rightarrow 0^{+}$ transitions)
nuclear $\beta$-decays, there are a number of corrections, some
nucleus dependent, that must be applied to the ft values. They are
collectively called RC in eq. (2). To make contact with previous
studies \cite{towner1, hardy}, we factorize them as follows:
\begin{equation}
1+RC= (1+\delta_{R})(1-\delta_{C})(1+\Delta)
\end{equation}
\noindent The first two factors are nucleus dependent while $\Delta$
is roughly nucleus independent, coming primarily from short-distance
loop effects. The axial-vector contributions discussed above are
included in the product $(1+\delta_{R})(1+\Delta)$. Because we include
leading logs from higher orders as well as some next-to-leading logs
\cite{marciano1, czarnecki}, the factorization is not exact and
$\Delta$ will exhibit some small nucleus dependence. In Table 1, we
give the decomposition of RC along with their uncertainties, using the
results from Hardy and Towner \cite{towner1, hardy} for nuclear
effects. The uncertainty in $1+\delta_{R}$ comes from
$Z^{2}\alpha^{3}$ and nuclear structure contributions while a common
$\pm 0.03 \%$ error in the Coulomb distortion effect is assigned to
$1-\delta_{C}$. The entire loop uncertainty decribed in this paper is
assigned to $1+\Delta$, even though much of it comes from medium and
long-distance effects and might better be attributed to
$1+\delta_{R}$. We have not made such a distribution, since it is
common to all decays.

Employing the corrections in Table 1 together with eqs. (2) and (18)
leads to the $V_{ud}$ values illustrated in Table 2. One finds for the
weighted average
\begin{equation}
V_{ud}= 0.97377(11)(15)(19) \ ({\rm Superallowed \ \beta-decays})
\end{equation}
\noindent Comparing with eq. (3) we see that our analysis gives a somewhat
larger $V_{ud}$ due to a $\pm0.00007$ increase from our new prescription along
with refinements from ref. \cite{czarnecki} which were not
included in Savard et al. \cite{savard}. Also, Savard et al. rounded
down in their analysis.

\newpage

\textbf{Table 1.} \ Decomposition of the RC for the nine best-measured
superallowed nuclear $\beta$-decays. Coulomb corrections in $1-\delta_{C}$
are taken directly from ref. \cite{towner1, hardy} while $1+\delta_{R}$ has
been somewhat modified due to our new results. The short-distance
$1+\Delta$ factor is based on the recent update in
ref. \cite{czarnecki} which includes higher order leading logs and
some next-to-leading logs.

\begin{center}
\begin{tabular}{c@{\hspace{2em}}c@{\hspace{1em}}c@{\hspace{1em}}c}
\hline
Nucleus & $1+ \delta_{R}$  &  $1- \delta_{C}$ & $1+\Delta$ \\
\hline
$^{10}$C & 1.01298(5)(35) & 0.9983(3) & 1.02389(38)\\
$^{14}$O & 1.01274(8)(50) & 0.9976(3) & 1.02385(38)\\
$^{26}$Al & 1.01468(21)(20) & 0.9971(3) & 1.02380(38)\\
$^{34}$Cl & 1.01343(34)(15) & 0.9939(3) & 1.02379(38)\\
$^{38}$K & 1.01322(41)(15) & 0.9939(3) & 1.02378(38)\\
$^{42}$Sc & 1.01469(49)(20) & 0.9954(3) & 1.02377(38)\\
$^{46}$V & 1.01392(57)(7) & 0.9959(3) & 1.02377(38)\\
$^{50}$Mn & 1.01394(65)(7) & 0.9957(3) & 1.02376(38)\\
$^{54}$Co & 1.01398(73)(7) & 0.9947(3) & 1.02376(38)\\
\hline
\end{tabular}
\end{center}
\newpage
\textbf{Table 2.} \ Values of $V_{ud}$ implied by various precisely
measured superallowed nuclear beta decays. The ft values are taken
from Savard et al. \cite{savard}. Uncertainties in $V_{ud}$ correspond
to: 1) nuclear structure and $Z^{2}\alpha^{3}$ uncertainties added in
quadrature with the ft error \cite{sirlin3, towner2}, 2) a common error assigned to nuclear
coulomb distortion effects \cite{towner2}, and 3) a common uncertainty from quantum
loop effects. Only the first error is used to obtain the weighted
average.

\begin{center}
\begin{tabular}{c@{\hspace{2em}} c@{\hspace{1em}} c@{\hspace{1em}} c}
\hline
Nucleus & ft(sec) & 1+RC & $V_{ud}$ \\
\hline
$^{10}$C & 3039.5(47) & 1.03542(36)(30)(38) & 0.97381(77)(15)(19)\\
$^{14}$0 & 3043.3(19) & 1.03441(52)(30)(38) & 0.97368(39)(15)(19)\\
$^{26}$Al & 3036.8(11) & 1.03582(30)(30)(38) & 0.97406(23)(15)(19)\\
$^{34}$Cl & 3050.0(12) & 1.03121(38)(30)(38) & 0.97412(26)(15)(19)\\
$^{38}$K & 3051.1(10) & 1.03099(44)(30)(38) & 0.97404(26)(15)(19)\\
$^{42}$Sc & 3046.8(12) & 1.03403(54)(30)(38) & 0.97330(32)(15)(19)\\
$^{46}$V & 3050.7(12) & 1.03376(59)(30)(38) & 0.97280(34)(15)(19)\\
$^{50}$Mn & 3045.8(16) & 1.03357(67)(30)(38) & 0.97367(41)(15)(19)\\
$^{54}$Co & 3048.4(11) & 1.03257(75)(30)(38) & 0.97373(40)(15)(19)\\
& & Weighted Average & 0.97377(11)(15)(19)\\
\hline
\end{tabular}
\end{center}
\newpage
We note that $^{46}V$ gives a somewhat low value for $V_{ud}$. It
differs from the average by 2.7 sigma. That particular nucleus
recently underwent a Q value revision \cite{savard} which lowered its
$V_{ud}$. It may be indicating problems with other Q values. If the
other nuclear Q values follow the lead of $^{46}V$, we could see a
fairly significant reduction in the weighted average for
$V_{ud}$. Clearly, remeasurements of Q values and half-lives of the
superallowed decays are highly warranted.

Employing the value of $V_{ud}$ in eq. (19) and the $K_{l3}$ average
\cite{blucher} for $V_{us}$
\begin{equation}
V_{us}=0.2257(9)(0.961/f_{+}(0)), \ K_{l3} \ {\rm average}
\end{equation}
with $f_{+}(0)=0.961(8)$ \cite{leutwyler} leads to the unitarity test 
\begin{equation}
|V_{ud}|^{2}+|V_{us}|^{2}+|V_{ub}|^{2}=0.9992(5)_{V_{ud}}(4)_{V_{us}}(8)_{f_{+}(0)}
\end{equation}

\noindent Good agreement with unitarity is found, with the dominant
uncertainty coming now from the theory error in the form factor
$f_{+}(0)$. Eq. (21) provides an important test of the standard model
at the quantum loop level and a constraint on new physics beyond the
standard model at the $\pm 0.09 \%$ level. We note, however, that some
other calculations \cite{blucher, jamin} of $f_{+}(0)$ and studies of
other strangeness changing decays \cite{marciano4} suggest a lower
$V_{us}$ value. Combined with further Q value revisions possibly
leading to a smaller $V_{ud}$, they could cause a significant
reduction in eq. (21). A future violation of unitarity is still
possible. However, for it to be significant, the theoretical
uncertainty in $f_{+}(0)$ must be further reduced.

\section {\bf Acknowledgments}
The work of W.J.M. was authored under Contract No. DE-AC02-98CH1086
with the U.S. Department of Energy.
The work of A.S. was supported by NSF Grant PHY-0245068.
Both authors would like to thank the Institute for Advanced Study for
support and its hospitality during the time this work was carried out.


\begin{thebibliography}{99}
\bibitem{towner1}
%\cite{Towner:2003cq}
%\bibitem{Towner:2003cq}
  I.~S.~Towner and J.~C.~Hardy, %``The Evaluation of V_{ud},
  Experiment and Theory,'' J.\ Phys.\ G {\bf 29}, 197 (2003)
  [arXiv:nucl-th/0308001]; %%CITATION = NUCL-TH 0308001;%%
%\cite{Wilkinson:1982hu}
%\bibitem{Wilkinson:1982hu}
  D.~H.~Wilkinson,
  %``Analysis Of Neutron Beta Decay,''
  Nucl.\ Phys.\ A {\bf 377}, 474 (1982);
  %%CITATION = NUPHA,A377,474;%%
%\cite{Pocanic:2003pf}
%\bibitem{Pocanic:2003pf}
  D.~Pocanic {\it et al.},
  %``Precise measurement of the pi+ $\to$ pi0 e+ nu branching ratio,''
  Phys.\ Rev.\ Lett.\  {\bf 93}, 181803 (2004)
  [arXiv:hep-ex/0312030].
  %%CITATION = HEP-EX 0312030;%%

\bibitem{bennett}
%\cite{Bennett:1999pd}
%\bibitem{Bennett:1999pd}
  S.~C.~Bennett and C.~E.~Wieman,
  %``Measurement of the 6S $\to$ 7S transition polarizability in atomic  cesium
  %and an improved test of the standard model,''
  Phys.\ Rev.\ Lett.\  {\bf 82}, 2484 (1999)
  [arXiv:hep-ex/9903022].
  %%CITATION = HEP-EX 9903022;%%

\bibitem{sirlin1}
%\cite{Sirlin:1977sv}
%\bibitem{Sirlin:1977sv}
  A.~Sirlin,
  %``Current Algebra Formulation Of Radiative Corrections In Gauge Theories And
  %The Universality Of The Weak Interactions,''
  Rev.\ Mod.\ Phys.\  {\bf 50}, 573 (1978)
  [Erratum-ibid.\  {\bf 50}, 905 (1978)];
  %%CITATION = RMPHA,50,573;%%
%\cite{Sirlin:1974ni}
%\bibitem{Sirlin:1974ni}
  A.~Sirlin,
  %``Radiative Corrections To G(V)/G(Mu) In Simple Extensions Of The SU(2) X
  %U(1) Gauge Model,''
  Nucl.\ Phys.\ B {\bf 71}, 29 (1974);
  %%CITATION = NUPHA,B71,29;%%
%\cite{Sirlin:1981ie}
%\bibitem{Sirlin:1981ie}
  A.~Sirlin,
  %``Large M (W), M (Z) Behavior Of The O (Alpha) Corrections To Semileptonic
  %Processes Mediated By W,''
  Nucl.\ Phys.\ B {\bf 196}, 83 (1982).
  %%CITATION = NUPHA,B196,83;%%

\bibitem{marciano1}
%\cite{Marciano:1985pd}
%\bibitem{Marciano:1985pd}
  W.~J.~Marciano and A.~Sirlin,
  %``Radiative Corrections To Beta Decay And The Possibility Of A Fourth
  %Generation,''
  Phys.\ Rev.\ Lett.\  {\bf 56}, 22 (1986).
  %%CITATION = PRLTA,56,22;%%

\bibitem{marciano2}
%\cite{Marciano:1982mm}
%\bibitem{Marciano:1982mm}
  W.~J.~Marciano and A.~Sirlin,
  %``Radiative Corrections To Atomic Parity Violation,''
  Phys.\ Rev.\ D {\bf 27}, 552 (1983);
  %%CITATION = PHRVA,D27,552;%%
%\cite{Marciano:1983ss}
%\bibitem{Marciano:1983ss}
  W.~J.~Marciano and A.~Sirlin,
  %``On Some General Properties Of The O (Alpha) Corrections To Parity Violation
  %In Atoms,''
  Phys.\ Rev.\ D {\bf 29}, 75 (1984)
  [Erratum-ibid.\ D {\bf 31}, 213 (1985)].
  %%CITATION = PHRVA,D29,75;%%

\bibitem{czarnecki}
%\cite{Czarnecki:2004cw}
%\bibitem{Czarnecki:2004cw}
  A.~Czarnecki, W.~J.~Marciano and A.~Sirlin,
  %``Precision measurements and CKM unitarity,''
  Phys.\ Rev.\ D {\bf 70}, 093006 (2004)
  [arXiv:hep-ph/0406324].
  %%CITATION = HEP-PH 0406324;%%

\bibitem{hardy}
%\cite{Hardy:2004dm}
%\bibitem{Hardy:2004dm}
  J.~C.~Hardy and I.~S.~Towner,
  %``New limit on fundamental weak-interaction parameters from superallowed beta
  %decay,''
  Phys.\ Rev.\ Lett.\  {\bf 94}, 092502 (2005)
  [arXiv:nucl-th/0412050];
  %%CITATION = NUCL-TH 0412050;%%
%\cite{Hardy:2004id}
%\bibitem{Hardy:2004id}
  J.~C.~Hardy and I.~S.~Towner,
  %``Superallowed 0+ $\to$ 0+ nuclear beta decays: A critical survey with tests
  %of CVC and the standard model,''
  Phys.\ Rev.\ C {\bf 71}, 055501 (2005)
  [arXiv:nucl-th/0412056].
  %%CITATION = NUCL-TH 0412056;%%

\bibitem{savard}
%\cite{Savard:2004fx}
%\bibitem{Savard:2004fx}
  G.~Savard {\it et al.},
  %``Q value of the superallowed decay of Mg-22 and the calibration of the
  %Na-21(p, gamma) experiment,''
  Phys.\ Rev.\ Lett. {\bf 95}, 102501 (2005).
  %%CITATION = PHRVA,C70,042501;%%

\bibitem{sirlin2}
A. ~Sirlin. Phys.\ Rev.\ {\bf 164},1767 (1967).

\bibitem{sirlin3}
%\cite{Sirlin:1987sy}
%\bibitem{Sirlin:1987sy}
  A.~Sirlin,
  %``Remarks Concerning The O (Z Alpha**2) Corrections To Fermi Decays,
  %Conserved Vector Current Predictions And Universality,''
  Phys.\ Rev.\ D {\bf 35}, 3423 (1987);
  %%CITATION = PHRVA,D35,3423;%%
%\cite{Jaus:1986te}
%\bibitem{Jaus:1986te}
  W.~Jaus and G.~Rasche,
  %``Radiative Corrections To O+ - O+ Beta Transitions,''
  Phys.\ Rev.\ D {\bf 35}, 3420 (1987);
  %%CITATION = PHRVA,D35,3420;%%
%\cite{Sirlin:1986cc}
%\bibitem{Sirlin:1986cc}
  A.~Sirlin and R.~Zucchini,
  %``Accurate Verification Of The Conserved Vector Current And Standard Model
  %Predictions,''
  Phys.\ Rev.\ Lett.\  {\bf 57}, 1994 (1986).
  %%CITATION = PRLTA,57,1994;%%

\bibitem{towner2}
%\cite{Towner:1994mw}
%\bibitem{Towner:1994mw}
  I.~S.~Towner,
  %``Quenching of spin operators in the calculation of radiative corrections for
  %nuclear beta decay,''
  Phys.\ Lett.\ B {\bf 333}, 13 (1994)
  [arXiv:nucl-th/9405031].
  %%CITATION = NUCL-TH 9405031;%%

%\cite{Towner:2002rg}
%\bibitem{Towner:2002rg}
  I.~S.~Towner and J.~C.~Hardy,
  %``Calculated corrections to superallowed Fermi beta decay: New evaluation of
  %the nuclear-structure-dependent terms,''
  Phys.\ Rev.\ C {\bf 66}, 035501 (2002)
  [arXiv:nucl-th/0209014].
  %%CITATION = NUCL-TH 0209014;%%

\bibitem{marciano3}
W. ~Marciano and A. ~Sirlin, a detailed account of our numerical and
analytical work will be presented in a later, longer manuscript.

\bibitem{bjorken}
%\cite{Bjorken:1966jh}
%\bibitem{Bjorken:1966jh}
  J.~D.~Bjorken,
  %``Applications Of The Chiral U(6) X (6) Algebra Of Current Densities,''
  Phys.\ Rev.\  {\bf 148}, 1467 (1966).
  %%CITATION = PHRVA,148,1467;%%

\bibitem{beg}
%\cite{Beg:1974vs}
%\bibitem{Beg:1974vs}
  M.~A.~B.~B{\'e}g,
  %``Anomalous Algebras And Neutrino Sum Rules,''
  Phys.\ Rev.\ D {\bf 11}, 1165 (1975);
  %%CITATION = PHRVA,D11,1165;%%
%\cite{Adler:1970hu}
%\bibitem{Adler:1970hu}
  S.~L.~Adler and W.~K.~Tung,
  %``Bjorken Limit In Perturbation Theory,''
  Phys.\ Rev.\ D {\bf 1}, 2846 (1970);
  %%CITATION = PHRVA,D1,2846;%%
%\cite{Bardeen:1978yd}
%\bibitem{Bardeen:1978yd}
  W.~A.~Bardeen, A.~J.~Buras, D.~W.~Duke and T.~Muta,
  %``Deep Inelastic Scattering Beyond The Leading Order In Asymptotically Free
  %Gauge Theories,''
  Phys.\ Rev.\ D {\bf 18}, 3998 (1978).
  %%CITATION = PHRVA,D18,3998;%%

\bibitem{gorishny}
%\cite{Gorishnii:1985xm}
%\bibitem{Gorishnii:1985xm}
  S.~G.~Gorishnii and S.~A.~Larin,
  %``QCD Corrections To The Parton Model Rules For Structure Functions Of Deep
  %Inelastic Scattering,''
  Phys.\ Lett.\ B {\bf 172}, 109 (1986);
  %%CITATION = PHLTA,B172,109;%%
%\cite{Larin:1991tj}
%\bibitem{Larin:1991tj}
  S.~A.~Larin and J.~A.~M.~Vermaseren,
  %``The alpha-s**3 corrections to the Bjorken sum rule for polarized
  %electroproduction and to the Gross-Llewellyn Smith sum rule,''
  Phys.\ Lett.\ B {\bf 259}, 345 (1991).
  %%CITATION = PHLTA,B259,345;%%

\bibitem{deur}
%\c%\cite{Deur:2004ti}
%\bibitem{Deur:2004ti}
  A.~Deur {\it et al.},
  %``Experimental determination of the evolution of the Bjorken integral at  low
  %Q**2,''
  Phys.\ Rev.\ Lett.\  {\bf 93}, 212001 (2004)
  [arXiv:hep-ex/0407007].
  %%CITATION = HEP-EX 0407007;%%


\bibitem{witten}
%\cite{Witten:1979kh}
%\bibitem{Witten:1979kh}
  E.~Witten,
  %``Baryons In The 1/N Expansion,''
  Nucl.\ Phys.\ B {\bf 160}, 57 (1979);
  %%CITATION = NUPHA,B160,57;%%

%\cite{Knecht:1999kg}
%\bibitem{Knecht:1999kg}
  M.~Knecht, S.~Peris and E.~de Rafael,
  %``A new approach to weak amplitudes in large N(C) QCD,''
  Nucl.\ Phys.\ Proc.\ Suppl.\  {\bf 86}, 279 (2000)
  [arXiv:hep-ph/9910396].
  %%CITATION = HEP-PH 9910396;%%

\bibitem{small}
Small $1/Q^{4}$ asymptotic contributions allowed by the data in
ref. (16) would have negligible effects on our analysis; see ref. (12)
for a detailed discussion.

\bibitem{blucher}
E. Blucher and W. Marciano to appear in PDG (2006).

\bibitem{leutwyler}
H.~Leutwyler and M.~Roos,
  %``Determination Of The Elements V (Us) And V (Ud) Of The Kobayashi-Maskawa
  %Matrix,''
  Z.\ Phys.\ C {\bf 25}, 91 (1984);
  %%CITATION = ZEPYA,C25,91;%%
D.~Becirevic et al., Nucl. \ Phys. \ B {\bf 705}, 339 (2005).

\bibitem{jamin}
%\cite{Jamin:2004re}
%\bibitem{Jamin:2004re}
  M.~Jamin, J.~A.~Oller and A.~Pich,
  %``Order p**6 chiral couplings from the scalar K pi form factor,''
  JHEP {\bf 0402}, 047 (2004)
  [arXiv:hep-ph/0401080];
  %%CITATION = HEP-PH 0401080;%%
%\cite{Bijnens:2003uy}
%\bibitem{Bijnens:2003uy}
  J.~Bijnens and P.~Talavera,
  %``K(l3) decays in chiral perturbation theory,''
  Nucl.\ Phys.\ B {\bf 669}, 341 (2003)
  [arXiv:hep-ph/0303103].
  %%CITATION = HEP-PH 0303103;%%

\bibitem{marciano4}
W.~J.~Marciano,
  %``Precise determination of $|$V(us)$|$ from lattice calculations of
  %pseudoscalar decay constants,''
  Phys.\ Rev.\ Lett.\  {\bf 93}, 231803 (2004)
  [arXiv:hep-ph/0402299];
  %%CITATION = HEP-PH 0402299;%%

 E.~Gamiz, M.~Jamin, A.~Pich, J.~Prades and F.~Schwab,
  %``V(us) and m(s) from hadronic tau decays,''
  Phys.\ Rev.\ Lett.\  {\bf 94}, 011803 (2005)
  [arXiv:hep-ph/0408044].
  %%CITATION = HEP-PH 0408044;%%




\end{thebibliography}
\end{document}